# Load Frequency Control of Solar PV and Solar Thermal Integrated Micro grid using Narma-L2 Controller

Sambit Dash

sambitdash.2011@gmail.com

**Abstract:** In this paper a novel approach to load frequency control of a microgrid with integrated Solar PV and Solar thermal generator is presented. A nonlinear adaptive controller based on an autoregressive moving average algorithm (NARMA) is used to control the frequency of the system. The objective of such a controller is to regulate and minimize frequency deviation due load changes. Further, a novel metaheuristic approach called Modified Whale Optimization Algorithm (MWOA) is used for creating an optimal PID controller to be used for frequency control. Finally, a comparative study between the proposed NARMA-L2 controller, WOA tuned PID controller and conventional PID controller is presented. From a series of experiments, it is concluded that the proposed NARMA-L2 controller outperforms other techniques of frequency control and gives the best result in terms of overshoot and settling time.

1. **Introduction:**

At the turn of the century, rapid industrialization and expansion caused a massive increase in electricity generation and consumption. With depleting conventional sources like coal and petroleum, research was the direction towards non conventional resources like wind, biomass, geothermal etc. In recent years PV systems have come to the forefront of such renewable sources. In an article by United Nations [1] it was mentioned that in rural homes of developing countries like India seven out of ten households use less efficient and harmful energy sources like kerosene along with the supply which is inconsistent and unreliable which makes these regions ideal for solar energy use. Although integration of solar energy to the grid helps to increase the amount of clean energy and promote ecological stability, but it is not without its own set of challenges. In a power system the load demand is not fixed also with the inclusion of Photo Voltaic and Solar thermal systems the generation component, which earlier used to be set the generation station is no longer fixed. The variation in both energy generation and load changes affects the frequency of the system drastically. Thus to keep the frequency under specified limits a load frequency controller is needed. Considering evolving energy sources, increasing size, development of new structures of power systems and increasing uncertainty in load and generation an optimal load frequency controller is one that could withstand sudden changes in load and maintain frequency deviation with specified range. There has been worldwide research in the field of frequency control. Over the years different techniques for controlling load frequency have been developed. A stand alone hybrid power system consisting of WTG, FC and AE were presented by Senjyu et al [2]. Asano et al [3] presented a paper on impact of Photo Voltaic power generation on Load frequency control. Daneshfar et al [4] proposed a Load frequency controller with genetic algorithm based multi objective control design. Bhatt et al [5] developed an optimized load frequency controller for multi area systems using particle swarm optimization. A combination of fuzzy logic and multiple tabu search optimization [6] based load frequency controller have also been developed. In a multi area system, Bacteria foraging based optimization is used to tune a fractional order PID [7]. Conventionally, PID is tuned heuristically but they do not offer optimal results thus non linear control strategies needed to tackle such scenarios. In recent years artificial neural networks are applied to nonlinear systems for identification and control. ANNs are capable of non linear mapping along with faster computational capabilities and implementation in parallel architecture. The main objective of this paper is the implementation of ANN based Narma L2 controller for load frequency control and comparison of this performance with modified whale optimization algorithm tuned PID controller and conventional PID controller.

## 2. System Modelling

The hybrid microgrid used in this paper consists of a conventional single machine system with multiple energy storage systems connected to improve efficiency and reliability of the model. Solar photo Voltaic and solar thermal power plant are also connected to the microgrid to provide load demand. A brief description of the components of the hybrid system is presented below:

### 2.1. Micro Turbine Generator

It is a high speed gas turbine with generator attached to it. Generally a permanent magnet synchronous generator is used. The turbine and generator are manufactured as single piece of rotating shaft system [8]. The small signal transfer function $G_{MTG}$ is denoted as $\frac{1}{T_{MTG}s+1}$ where the value of $T_{MTG} = 2$.

### 2.2. Diesel Engine Generator

Automatic start and supply DEG are used to provide deficit energy incase of energy shortfall. DEG is denoted by a linear first order transfer function with $T_{DEG}$ as the time constant. $G_{DEG} = \frac{1}{T_{DEG}s+1}$

## 2.3. Fuel Cell

Fuel Cell is a electrochemical energy conversion device where chemical energy is directly converted to electrical energy. Fuel cells are static energy conversion devices. They provide considerable efficiency and reliability in a microgrid. The linear model of fuel cell is represented by a simple first order transfer function. $G_{FC} = \frac{1}{T_{FC}s+1}$ where $T_{FC} = 4$.

## 2.4. Flywheel Energy Storage System

In FESS, kinetic energy is stored with the help of flywheel inertia. During off peak conditions electrical energy is used to run the flywheel and during peak condition energy is generated using the rotating flywheel. Generally FESS are fast acting devices and smooth out minor load disturbance with considerable ease. The linear first order transfer function used in FESS is denoted by $G_{FESS} = \frac{1}{T_{FESS}s+1}$ and $T_{FESS}$ denoted the time constant of flywheel where value of $T_{FESS} = 0.1$

## 2.5. Battery Energy Storage System

In BESS, during charging energy is stored in the form of chemical ions and when connected to a load electrical energy is discharged. With the advancements in power electronics, batteries are used in conjunction with power converters to change DC to AC power and to damp out harmonics that might be generated. The transfer function denoting BESS is given as $G_{BESS} = \frac{1}{T_{BESS}s+1}$ where the value of $T_{BESS} = 0.1$

## 2.6. Photovoltaic Power Generation

A solar PV plant consist of multiple PV panels and each panel consist of several solar cells. Thus solar cell is the fundamental component of Solar power generation. The electrical model of a practical solar cell consist of a current source, a diode in parallel with shunt and series resistance. Observing the equivalent circuit and applying KCL the net current produced by solar cell is

$I = I_L - I_D - I_{SH}$ where $I_L$ is the photo electric current, $I_D$ is the diode current, $I_{SH}$ is the current through the shunt resistance and $I$ is the net current output of solar cell.

Further the diode current $I_D$ is given by Shockley equation

$$I_D = I_0 \left\{ \exp\left[\frac{V_j}{nV_T}\right] - 1 \right\}$$

where $I_0$ is the reverse saturation current, $V_T$ is the thermal voltage, $V_j$ is the voltage across diode and resistor and n is the diode ideality factor.

The shunt resistance and the Voltage across diode are governed by following equation

$$I_{SH} = \frac{V_j}{R_{SH}} \text{ and } V_j = V + IR_s$$

Thus the net current output of solar cell can be denoted by

$$I = I_L - I_0 \left\{ \exp\left[\frac{V+IR_s}{nV_T}\right] - 1 \right\} - \frac{V+IR_s}{R_{SH}}$$

The transfer function used to denote the PV plant

$G_{PV} = \frac{1}{T_{PV}s+1}$ where $T_{PV} = 1.8$

## 2.7. Solar Thermal power generation

These plants basically use reflecting mirrors or collector to focus the sun's rays at a single point. This accumulation of sun's rays increase the temperature of the collector which is then used to boil water to generate steam. This super heated steam is used to run the turbine. The transfer function of STPG plant is denoted as $G_{STPG} = \frac{1}{T_{STPG}s+1}$ and $T_{STPG} = 1.2$

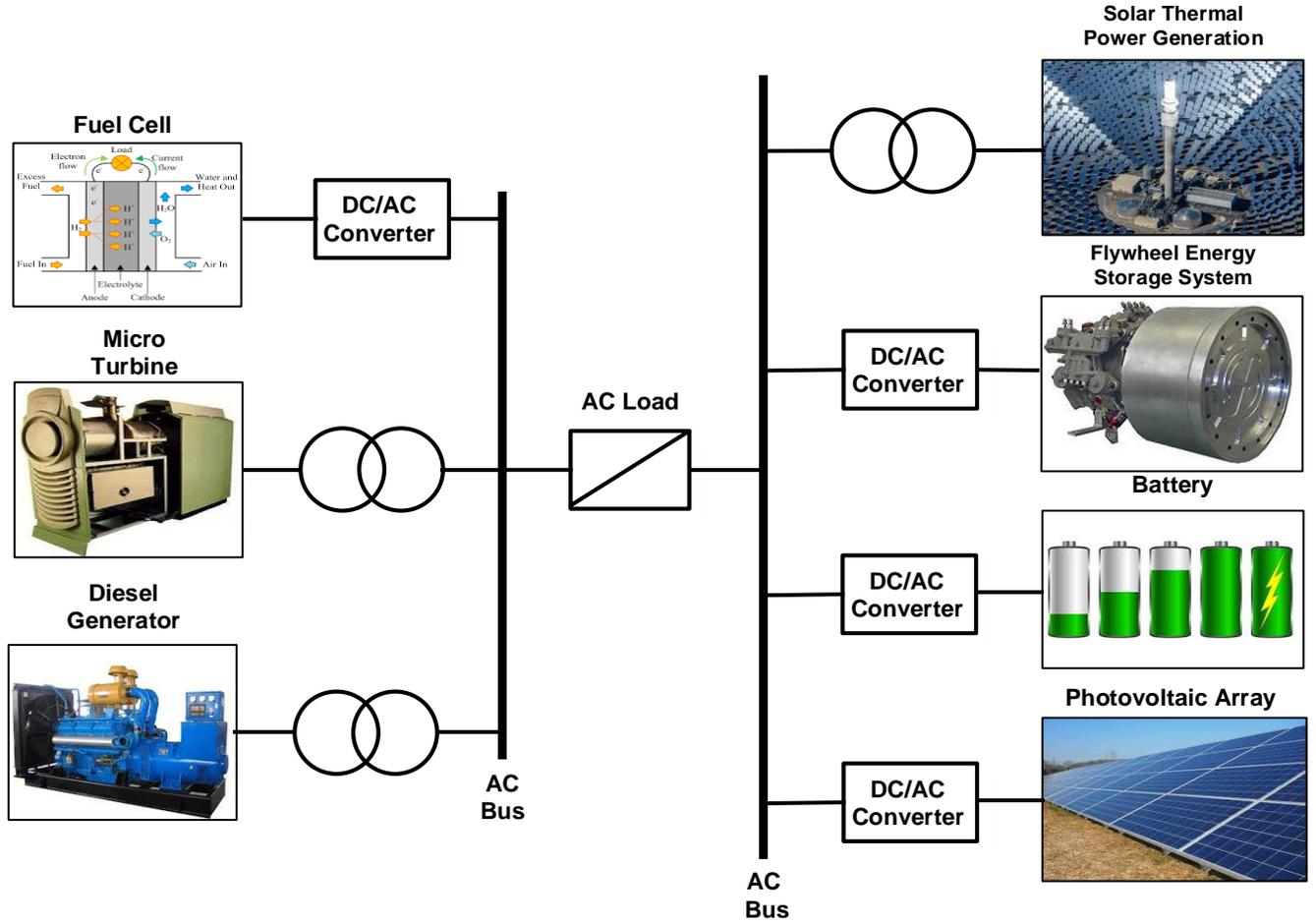

Fig. 1: A simple scheme of PV and Solar thermal integrated microgrid

Using power electronics converters the Distributed sources are connected to the AC bus. The equation (9-11) representing the net power generation for supplying the load is given by

$$P_{Load} = P_{DEG} + P_{MTG} + P_{PV} + P_{STPG} + P_{FC} \pm P_{BES} \pm P_{FES}$$

The power generated by solar PV and Solar thermal are stochastic in nature but for simpler analysis they are taken as constant. When a step load is applied the Renewable energy sources i.e PV and STPG provide power along with the energy storage devices such that both the demand and supply sides are balanced.

$$\Delta P_{Load} + \Delta P_{DEG} + \Delta P_{MTG} + \Delta P_{PV} + \Delta P_{STPG} + \Delta P_{FC} + \Delta P_{BES} + \Delta P_{FES} = 0$$

**Parameters of different components of microgrid:**

| Parameter | Value | Parameter | Value |
|---|---|---|---|
| D(pu/Hz) | 0.012 | $T_{DEG}$ | 2 |
| M(pu/s) | 0.2 | $T_{MT}$ | 2 |
| $T_{FC}$ | 4 | $T_{PV}$ | 1.8 |
| $T_{BESS}$ | 0.1 | $T_{STPG}$ | 1.2 |
| $T_{FESS}$ | 0.1 | | |

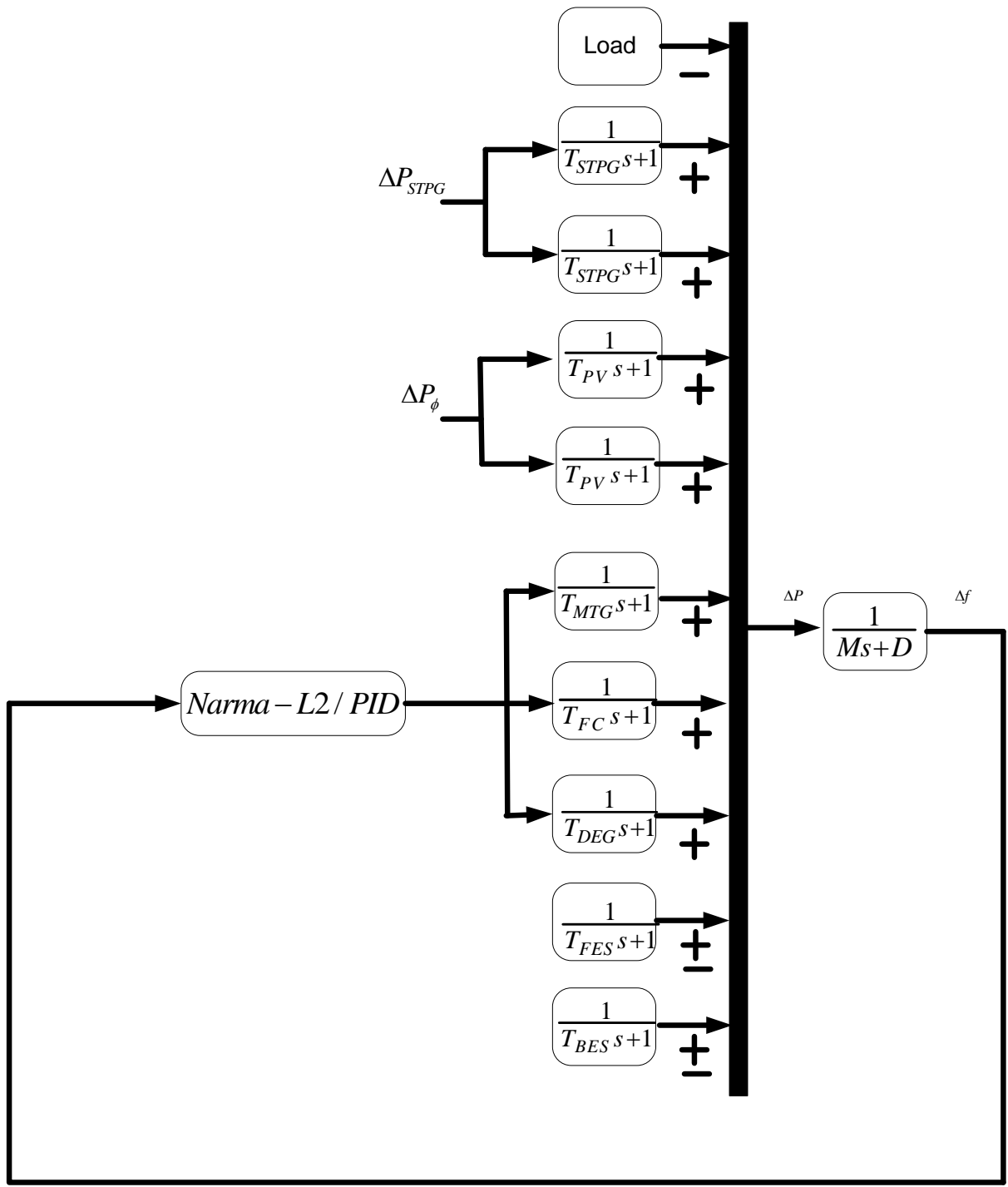

Fig. 2. Dynamic frequency response model of microgrid

## 3. Controller Structure

### 3.1. Modified Whale optimization Algorithm (MWOA)

Whale optimization algorithm is a novel metaheuristic technique based on the behavior of whales [12]-[13].This optimization technique is based on humpback whales. These whales feed on Krill and small fishing herds. A bubblenet feeding approach is adopted by humpback whales while hunting. In WOA ,the mechanism to find the best solution consist of making it a target to other search agents [14].Thus to nullify such action the prey changes position in accordance with best search agent. Since the region containing the optimal minima is not known beforehand this may lead to the solution being trapped in local minima. The position vectors change during search and if the procedure is filled with large steps this may result in improper searching and inability to find the local optima. To minimize the steps, correction factors were introduced in modified whale optimization technique (MWOA).The equation thus becomes

$$\vec{D} = \left| \vec{C}.\vec{X}^*(t) - \vec{X}(t) \right| \Big/ CF_1$$

$$\vec{X}(t+1) = (\vec{X}^*(t) - \vec{A}.\vec{D}) \Big/ CF_1$$

With the introduction of search space the whales move towards the prey in small steps thereby exploring the search area more efficiently.

Similarly in the exploitation phase also the correction factor term is introduced where the spiral updation is given by

$$\vec{X}(t+1) = (\vec{D}'.e^{bl}.\cos(2\pi l) + \vec{X}^*(t)) \Big/ CF_2$$

Due to the presence of Correction factor $CF_2$, the whales swim in a small circular area which increases the exploitation capability of the algorithm. Finally in the exploration phase, correction factor is also added.

$$\vec{D} = \left| \vec{C}.\vec{X_{rand}} - \vec{X} \right| \Big/ CF_1$$

$$\vec{X}(t+1) = (\vec{X_{rand}} - \vec{A}.\vec{D}) \Big/ CF_1$$

Thus by introducing correction factor and randomness the capability of the whales reaching any position in the search space is enhanced. The optimal values found using MWOA algorithm is used in PID block and simulation is conducted for the given microgrid.

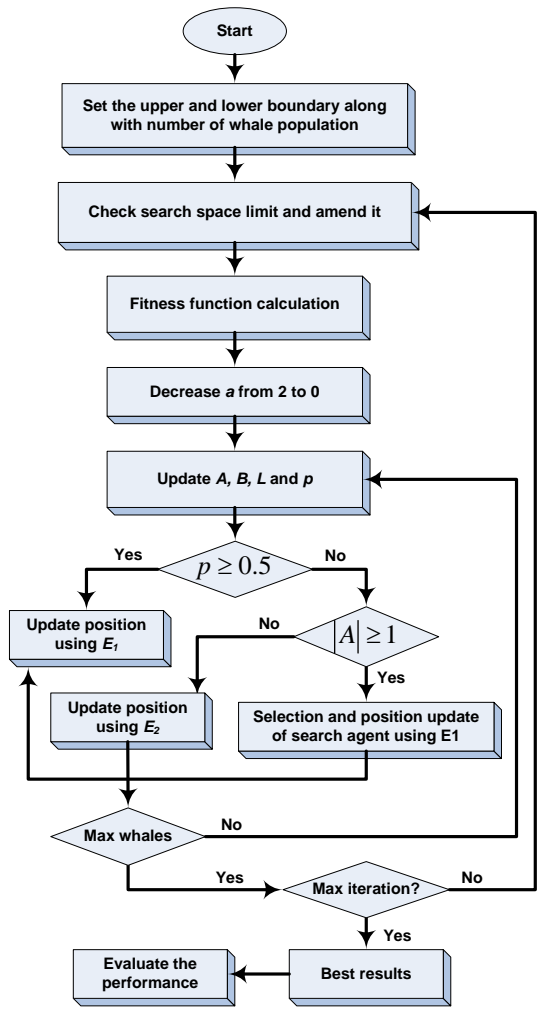

### 3.2. Narma L2 controller:

Non Linear Autoregressive Moving Average (NARMA) model is the most discernible form of non linear discrete time system. In a NARMA model the past, present input and output values are used to determine future output values.

$$y_{(k+d)} = f[y_{(t)}, y_{(t-1)}, \ldots y_{(t-n+1)}, u_{(t)}, u_{(t-1)}, \ldots u_{(t-n+1)}]$$ where y(t) denotes the output and u(t) denotes the input and f() denotes the non linear approximation between y(t) and u(t). For control theory application, back propagation based ANN proved to be too slow thus an efficient method involving approximate model was introduced by Mukhopadhyay et al [15]. Two methods of NARMA have been proposed; NARMA L1 and NARMA L2. From mathematical standpoint it can be known that NARMA-L1 need n+1 neurons to compute whereas the NARMA-L2 requires just 2 neurons to compute. This makes NARMA-L2 not only superior but also optimal for practical applications. NARMA L2 involves two sub approximation functions which are efficient in adaptation of control context.

$$y^* = f[y_{(t)}, y_{(t-1)}, \ldots y_{(t-n+1)}, u_{(t)}, u_{(t-1)}, \ldots u_{(t-n+1)}] + g[y_{(t)}, y_{(t-1)}, \ldots y_{(t-n+1)}, u_{(t)}, u_{(t-1)}, \ldots u_{(t-n+1)}] \times u_{(t)}$$

The functions f and g are also used in control phase

$$u_{(t)} = \frac{y_{(k+d)} - f[y_{(t)}, y_{(t-1)}, \ldots y_{(t-n+1)}, u_{(t)}, u_{(t-1)}, \ldots u_{(t-n+1)}]}{g[y_{(t)}, y_{(t-1)}, \ldots y_{(t-n+1)}, u_{(t)}, u_{(t-1)}, \ldots u_{(t-n+1)}]}$$

The main concept of this controller is shifting of non linear system dynamic to linear system by removal of nonlinearities in mapping. In Narma-L2, a simple rearrangement of the neural network of the plant to be controlled, which is trained offline, in batch form [16]-[19].The block diagram of Narma L2 controller consist of 2 multilayer neural networks namely f and g. Two tapped delay lines present in controller structure are used to store past values of input and output signal. Basically Narma-L2 controller first uses random input which is fed into the plant and stores corresponding output. Then it maps the relationship between the input and output using a linear structure consisting of weights and minimizes the weight equation iteratively until the optimal weight which best describes correlates the input and output is achieved. Using the same weight finally the controller is tested on new input values. In this paper we use the Narma L2 controller provided in the SIMULINK library of MATLAB software.

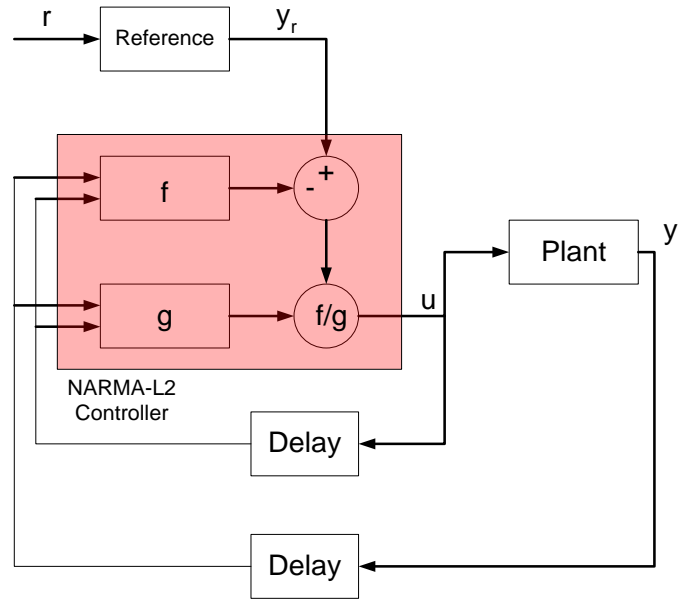

Fig. 3.A simplified diagram of NARMA-L2 controller

## 4. Result Analysis

Here a photovoltaic and solar thermal integrated microgrid incorporated with Narma-L2 controller is used to minimize the change in frequency and damping perturbation produced due to load changes. Further the response of Narma-L2 controller is compared with Modified whale optimization algorithm (MWOA) tuned PID controller. The use of Narma –L2 controller consists of 3 phases. Initially the controller is trained where the response of the plant for different random input is collected. The artificial neural network of the controller is trained using levenberg-Marquardt algorithm. The hidden layer of ANN consists of 10 neurons and 10000 sample points are used to train the network. The figure 1,2,3    show the training, validation and testing phase of the network and the plant output, neural network output, input, error in each case. Finally the controller is tested on the actual hybrid power system. A comparison of frequency deviation between Narma-L2, MWOA PID and conventional PID for a load of 0.2 p.u is shown in figure 7,8 .The figure  show the response of BESS,FC and MTG for the given load condition. It can be observed that Narma-L2 controller gives superior results both in terms of overshoot and settling time than MWOA tuned PID and conventional PID.

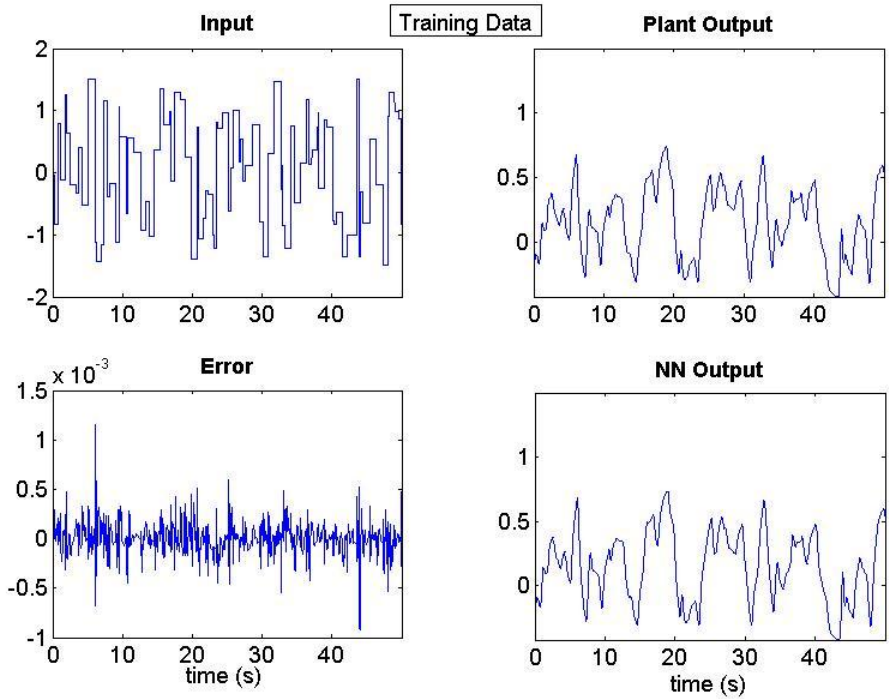

Fig. 4. Training Data of Narma-L2 controller

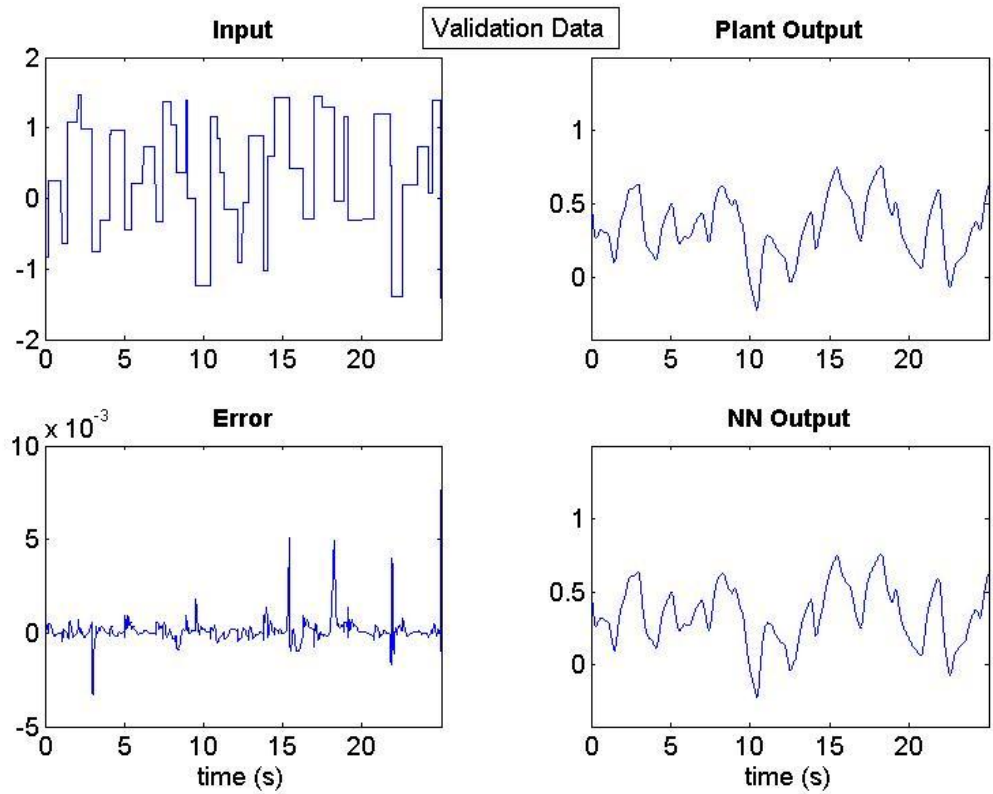

Fig.5. Validation data of Narma-L2 controller

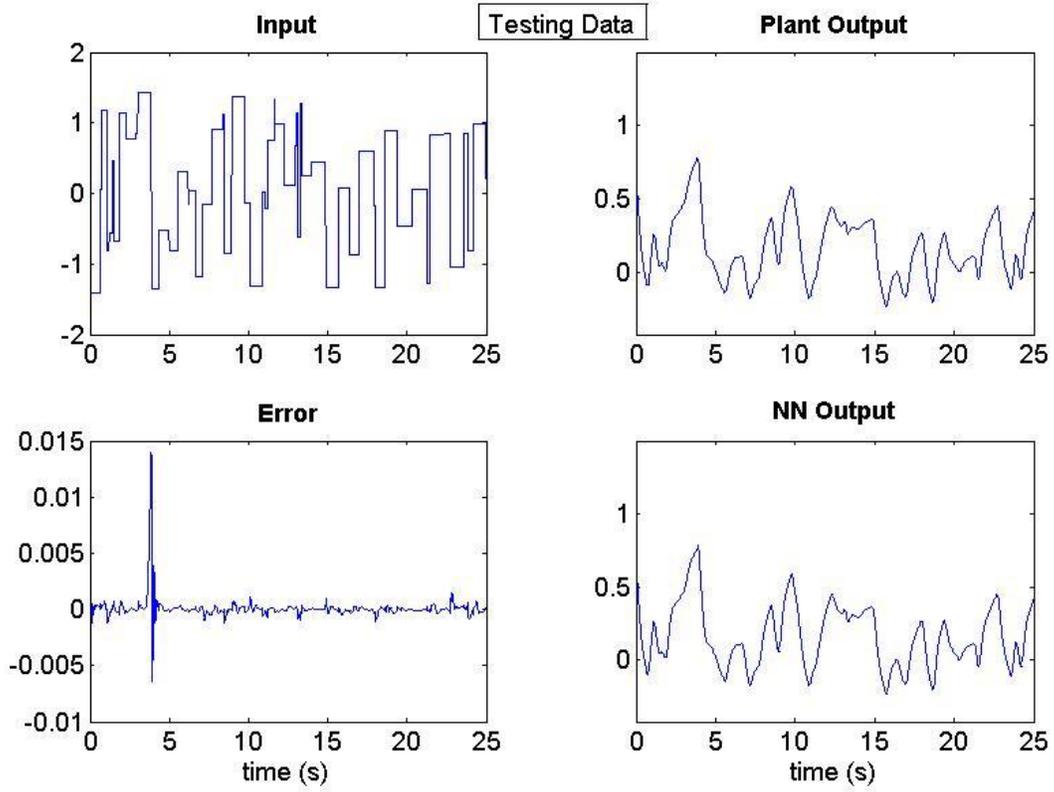

Fig 6. Test Data of Narma-L2 controller

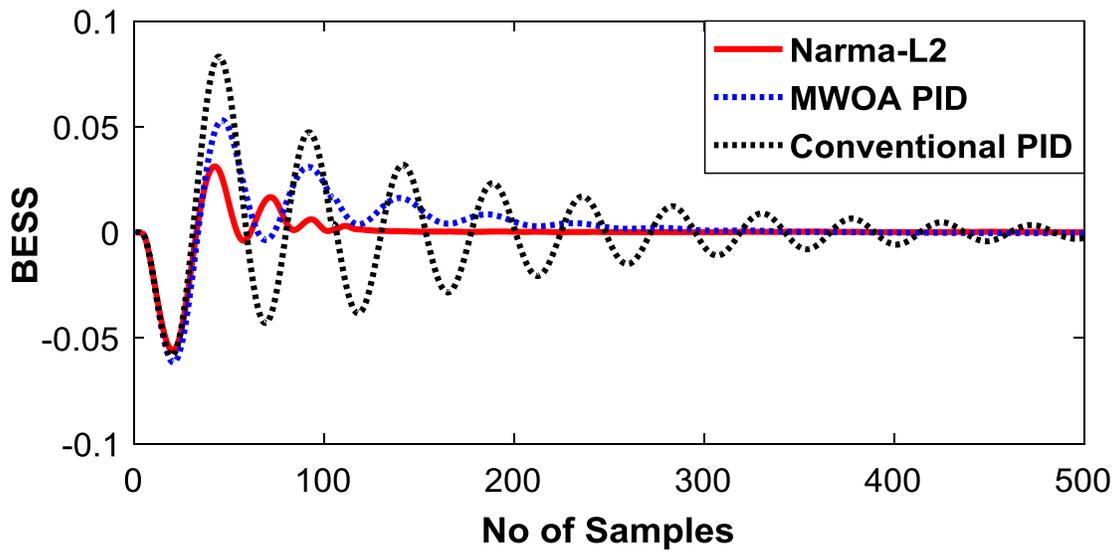

Fig. 7. Output power of BESS for different controllers

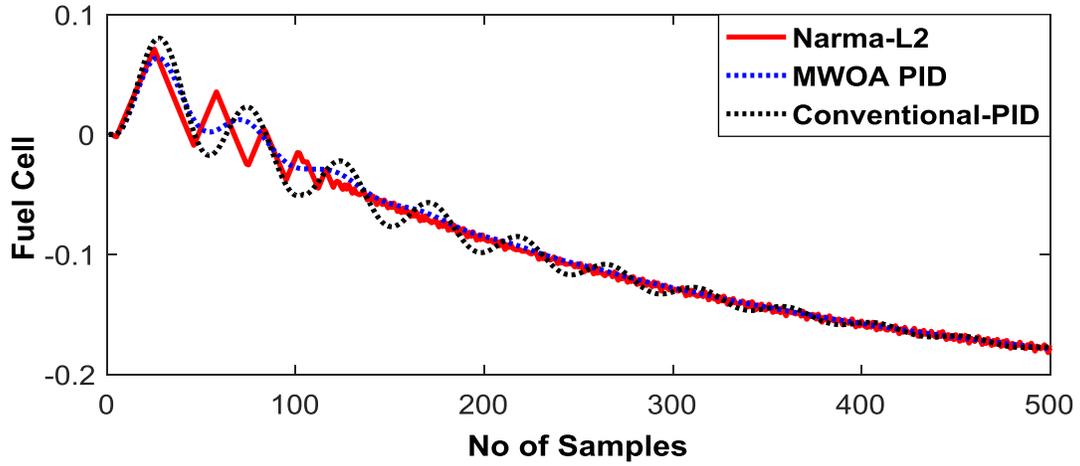

Fig. 8. Output of FC for different controllers

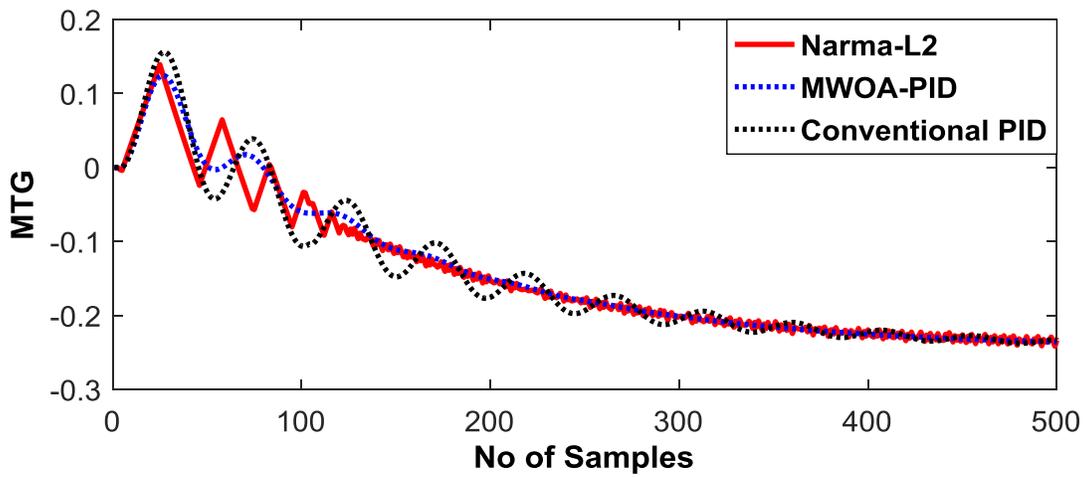

Fig. 9. Output of MTG for different controllers

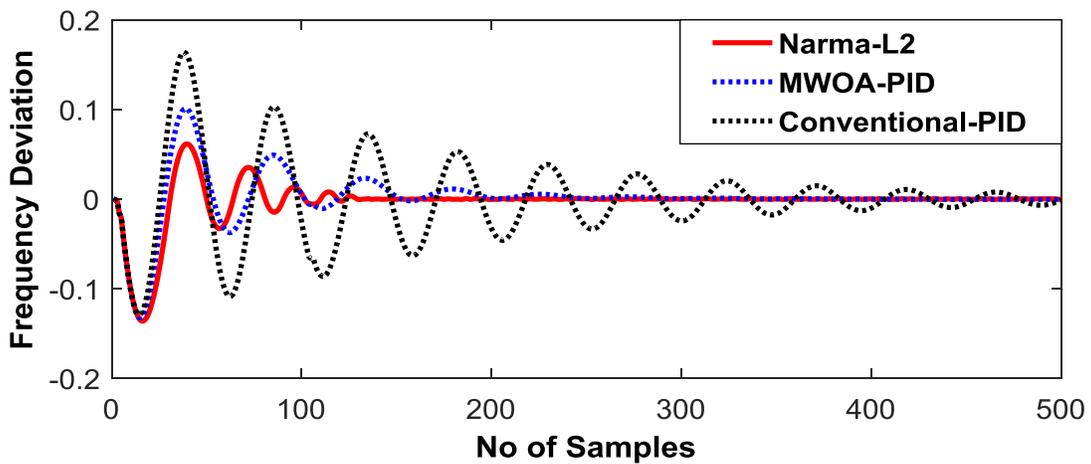

Fig. 10. Frequency deviation of for a load of 0.3 pu for different controllers

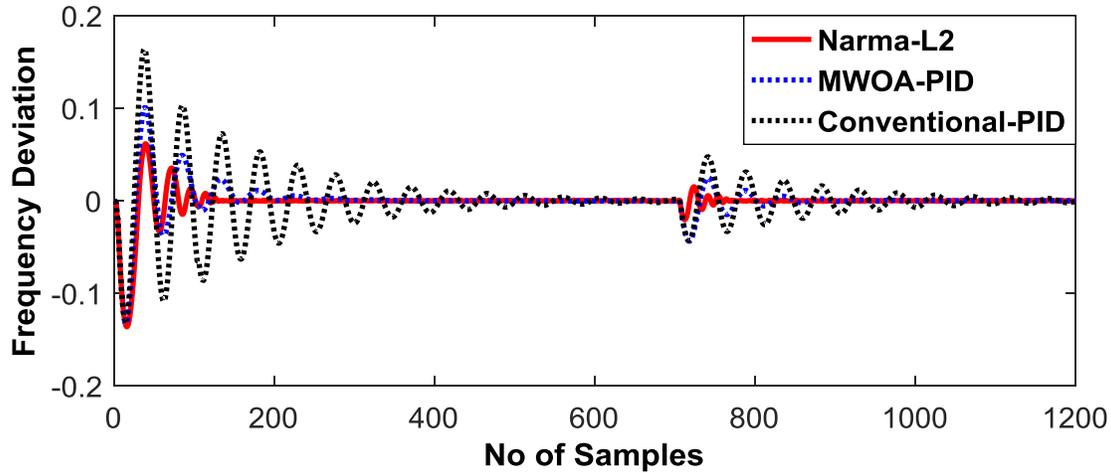

Fig 11. Frequency deviation for load of 0.4 + 0.3 u(t-4).

## 5. Conclusion

In this paper a Photovoltaic and Solar thermal based microgrid is simulated for different loads. The frequency control of the hybrid system using Narma-L2 controller, MWOA tuned PID and conventional PID is simulated. Training ,Validation and testing data of the Narma-L2 controller along with the plant , neural network output and their corresponding error in estimation is plotted. The MWOA algorithm is iterated multiple times to find the optimal minima.. The response of different energy storage elements such as BESS, FC and MTG under different load and different controllers is recorded. It is observed that Narma-L2 controller gives the best response in reducing frequency deviation and overshoot as compared to other controllers under various conditions.